\documentclass[10pt, journal, twocolumn, letterpaper]{IEEEtran}
\usepackage{amsmath,amssymb}
\usepackage{graphicx,epsfig}
\newcommand{\I}{{1\!\!1}}

\newcommand{\R}{{\mathbb{R}}}

\def\vs{\vspace{.3cm}}

\newcommand{\beq}{\begin{equation}}
\newcommand{\eeq}{\end{equation}}

\usepackage{bm}

\def\YY{\mathcal{Y}}
\def\G{\mathcal{J}}
\def\I{\mathcal{I}}
\def\RR{\mathcal{R}}
\def\c{c}
\def\R{\mathcal{L}}
\def\S{\mathcal{\omega}}
\def\D{\mathcal{D}}
\def\ee{\end{equation}}
\def\bea{\begin{eqnarray}}
\def\eea{\end{eqnarray}}

\def\OO{\mathcal{R}}
\def\VV{\mathcal{V}}
\def\opt{}

\def\part{{\cal P}_{[s]}}

\def\CK{Csisz\'ar and K\"orner }

\newcommand{\proj}[1]{\,\mathbf{d}_{#1}\,\mathbf{d}^{\dagger}_{#1}}
\newcommand{\w}[1]{(\ref{#1})}

\title{Key Distillation and the Secret-Bit Fraction}

\author{Nick S. Jones$^\dag$\thanks{$^\dag$OCISB, Department of Physics, University of Oxford, Parks
Road, Oxford. OX1 3PU, $^\ddag$Department of Applied Mathematics and
Theoretical Physics, University of Cambridge, Wilberforce Road,
Cambridge CB3 0WA, U.K.} and Llu\'{\i}s Masanes$^\ddag$}

\begin{document}

\maketitle

\begin{abstract}

We consider distillation of secret bits
from partially
secret noisy correlations $P_{ABE}$, shared between
two honest
parties and an eavesdropper. The most studied distillation scenario
consists of joint operations on a large
number of copies of the distribution $(P_{ABE})^N$,
assisted with
public communication. Here we consider distillation
with only
one copy of the distribution, and instead of rates,
the `quality' of
the distilled secret bits is optimized, where the
`quality' is
quantified by the secret-bit fraction of the result.
The secret bit
fraction of a binary distribution is the
proportion which constitutes a secret
bit between
Alice and Bob. With local operations and public
communication
the maximal extractable secret-bit fraction from a
distribution
$P_{ABE}$ is found, and is denoted by
$\Lambda[P_{ABE}]$. This quantity is shown to be
nonincreasing under local operations and
public communication, and nondecreasing under
eavesdropper's local operations:  $\Lambda$
is a secrecy monotone. It is shown that if
$\Lambda[P_{ABE}]> 1/2$
then $P_{ABE}$ is distillable, thus providing a
sufficient condition
for distillability. A simple expression for
$\Lambda[P_{ABE}]$ is
found when the eavesdropper is decoupled, and
when the honest
parties' information is binary and the local operations are
reversible. Intriguingly, for
general distributions the (optimal) operation requires local
 degradation of the
data.

\end{abstract}


\section{Introduction}

If two parties are to communicate with perfect secrecy over an
insecure channel, they must share a secret key at least as long as
the message to be transmitted \cite{Shannon,note2}. It is, however,
not always necessary for the two parties (Alice (A) and Bob (B)) to
meet up in order to obtain a shared secret key
\cite{Wyner,CK,maurer}. It might be the case that, secret key aside,
the three parties (Alice, Bob and Eve (E) the eavesdropper) have
access to an information source which provides partially correlated
data to each of them. These correlations can be captured by a
tripartite probability distribution $P_{ABE}$. If Eve has access to
the same information as Alice and Bob, secure key generation is
impossible. However, there are many possible physical scenarios in
which this perfect correlation is not present; in these cases this
difference in knowledge can sometimes be exploited to generate
secret key.

\vs Inspired by closely related work by Wyner, and \CK
\cite{Wyner,CK}, Maurer \cite{maurer} presented a protocol for
secret key agreement by public discussion which exploits such
imperfect knowledge. In his approach Alice and Bob are given access
to an insecure, authenticated, tamper-proof channel and also receive
sample data from a distribution $P_{ABE}$. In an example, he
considers the distribution  generated when a satellite broadcasts
the same random bits to each party but Alice, Bob and Eve receive
the information down binary symmetric channels with bit errors of
$20\%$, $20\%$ and $15\%$ respectively. Even though Eve's error is
less than Alice or Bob's, Maurer provides a procedure, called
advantage distillation, which allows them to obtain shared random
bits about which Eve knows arbitrarily little. Maurer, with Wolf,
subsequently provided an if and only if distilability condition for
all distributions created by a combination of a satellite producing
random bits and local noise \cite{MW1,MW}.

\vs Note that it is assumed that all parties know the distribution
$P_{ABE}$. The knowledge they lack is only about particular samples
from the distribution. We will also be making this assumption
throughout the following. This is not an innocent postulate; though
it is sensible to assume that Eve knows $P_{ABE}$, one need not
assume that Alice and Bob know anything about Eve's data. Advantage
distillation requires that Alice and Bob have a bound on Eve's error
rate. If the physical situation prevents them bounding her errors,
the parties might be better off using quantum cryptography
\cite{GisincryptoRMP}.

\vs If Alice and Bob want to communicate secretly, they will not
always have a satellite available to help them generate their
secret key. The broad question addressed in this paper is then:
what physical situations can be used to generate secret key? Or
more precisely, which distributions, $P_{ABE}$, can be used to
generate secret key?

\vs The approach in this paper is rather different from that adopted
in other work (though it is related to a construction in
\cite{Renner} and in 
\cite{Holenstein05} see Section \ref{secIIkey}). In the usual
scenario the distillation procedure consists of joint operations on
an arbitrarily large number of copies of the distribution
$(P_{ABE})^N$, assisted by communication over an insecure, but
authenticated channel. In this context, the secrecy properties of a
distribution $P_{ABE}$ are typically assessed by the `secret key
rate'. This is the maximal rate at which Alice and Bob, receiving
data according to $P_{ABE}$, can generate a key about which Eve's
information is arbitrarily close to zero. By contrast, we consider
distillation in the `single-copy' scenario, and instead of rates the
protocol optimizes the `quality' of the distilled secret bit, where
the `quality' is quantified by the secret-bit fraction of the
result. The secret bit fraction of
$P_{ABE}$ is defined as 
the maximum $\tau$ such that there exists a decomposition of ${
P}_{ABE}$ of the form: $ {  P}_{ABE}=\tau {  S}_{AB}{  Q}_E +
(1-\tau){  H}_{ABE}$ where $\tau\in[0,1]$, ${  Q}_E$ and ${
H}_{ABE}$ can be any probability distributions and ${ S}_{AB}$ is a
shared bit.

\vs Given a distribution $P_{ABE}$, the maximal `quality' of the
secret bits that can be distilled from it is denoted by
$\Lambda[P_{ABE}]$, and called the `maximal extractable secret-bit
fraction' (MESBF) of $P_{ABE}$.

\vs We define $\Lambda[P_{ABE}]$ as follows. Suppose Alice, Bob and
Eve all receive one sample from the distribution $P_{ABE}$. Consider
the set of distributions $P'_{ABE}$ that can be obtained from
$P_{ABE}$ with some probability, when Alice and Bob perform local
operations and public communication (LOPC). We allow the probability
of obtaining any such $P'_{ABE}$ to be arbitrarily small as long as
it is positive. We call this class of transformations
stochastic-LOPC (SLOPC). We also call them filtrations or filtering operations. We consider SLOPC transformations because,
as mentioned above, we do not care about the rates at which the
distributions $P'_{ABE}$ can be obtained from $P_{ABE}$. Instead, we
want to know which of the obtainable distributions $P'_{ABE}$ most
resembles   a secret bit, and we quantify this resemblance by the
secret-bit fraction. We denote the maximal secret-bit fraction that
can be extracted from $P_{ABE}$ by $\Lambda[P_{ABE}]$.

\vs If Alice and Bob share a perfectly correlated random bit and Eve
is uncorrelated from them,  $\Lambda[P_{ABE}]$ will be `1'. If all
parties only have uncorrelated data as outputs then
$\Lambda[P_{ABE}] =1/2$. Note that the filtrations can sometimes
fail. This failure rate is not reflected in the size of
$\Lambda[P_{ABE}]$ since we only consider the case where the
filtration is successful. It follows that distributions exist with
$\Lambda[P_{ABE}]$ equal to `1' but with very low secret key rates.

\vs One of the main results motivating our use of the MESBF is to
show that if $\Lambda[P_{ABE}]>\frac{1}{2}$ then $P_{ABE}$ has a
positive secret key rate (in the asymptotic scenario). The value of
$\Lambda[P_{ABE}]$ can thus be an indicator of whether a
distribution has distillable key: however it tells us nothing about
the size of the secret key rate. A
necessary and sufficient condition for distributions to 
have secret key is that there exists a positive integer $N$ such
that $\Lambda[P^N_{ABE}]>\frac{1}{2}$, where $P^N_{ABE}$ represents
$N$ samples from $P_{ABE}$ \cite{note1}.

\vs A very similar quantity called the `singlet-fraction' has been
introduced in entanglement theory in quantum mechanics, in the
context of entanglement distillation \cite{horosinglet}. To our
surprise we were able to prove rather more about our classical
quantity that has been found for the quantum case. The connection
between entanglement theory and cryptography is not coincidental and
has been investigated at length (one of the best introductions is
\cite{CP}).  In analogy to bound entanglement \cite{horobound1}, the
existence of bound information has been conjectured
\cite{GW,Renner03}. Distributions that can yield no secret key and
yet cannot be created by LOPC show bound information. A distribution
will have bound information if $\Lambda[P^N_{ABE}]=\frac{1}{2}$ for
all $N$ and yet the distribution cannot be generated by LOPC alone.
Hence, the study of $\Lambda$ may prove useful for proving the
existence of bound information.

\vs Let us now highlight the results in this paper. As well as
showing (a) that $\Lambda[P_{ABE}]>\frac{1}{2}$ implies a positive
secret key rate we present four further results. (b) We show that
$\Lambda[P_{ABE}]$ is a secrecy monotone under SLOPC by Alice and
Bob and under local operations by Eve. (c) We have a closed
expression for $\Lambda[P_{ABE}]$ for all distributions where Eve is
uncoupled, that is $P_{ABE}=P_{AB}P_E$. In this case, the optimal
filtration is also obtained.
(d) We find $\Lambda[P_{ABE}]$ for $P_{ABE}$ where Alice and Bob's
random variables only have two possible outcomes and are restricted
to using filtrations which can be stochastically reversed. (e) We
show that, for general $P_{ABE}$,  optimal filtering operations can
sometimes require Alice and Bob to degrade their data (by partially
locally randomizing). This last result is surprising. One might
expect that if Alice and Bob degrade their information they will
have a lower secret-bit fraction; however this is to neglect the
role of Eve who might lose, comparatively, even more information. We
provide an example where local randomization improves the secret-bit
fraction of a distribution over that obtained when the data is
reversibly transformed.

\vs A brief outline of the rest of this paper is now given.
Section \ref{secIkey}
 introduces the scenario considered, defines the notation, and presents the first results including the proof that $\Lambda[P_{ABE}]$ is a secrecy monotone.
 Section
\ref{secIIkey} supplies a sufficient condition for a distribution to
be used to generate secret key. Section \ref{secIIIkey} describes
reversible filtrations, operations which can be successfully undone
with a non-zero probability. The same section finds
$\Lambda[P_{ABE}]$ for distributions where Alice and Bob can only
have two outcomes and perform reversible filtrations. Section
\ref{secVkey} finds $\Lambda[P_{AB}]$ when Eve is decoupled from the
communicating parties. The last section of results, \ref{secVIkey},
shows that in general, filtrations that yield the MESBF require the
cooperating parties to degrade their data. We conclude by discussing
open problems and investigating interpretations of the quantity
$\Lambda[P_{ABE}]$. The appendices contain some of the longer
proofs; Appendix \ref{decompsec} is of independent interest as it
provides a useful general decomposition of filtrations.

\section{Definitions and basic results}\label{secIkey}

In the following we define the scenario considered in this paper.
Alice and Bob are connected by an authenticated tamper-proof
channel. The channel is, however, insecure; a third party, Eve,
learns all communicated messages. Alice, Bob and Eve each obtain a
letter from  alphabets of sizes $d_A,d_B,$ and $d_E$ respectively.
These outputs come from a probability distribution ${ P}_{ABE}$.
Here, and in what follows, $A, B, E$ will only appear as labels
identifying the parties sampling from the distribution ($A, B, E$
are not random variables). The symbols $a,b,e$ will be treated as
random variables with alphabets of size $d_A,d_B,$ and $d_E$
respectively. The same symbols $a,b,e$ will also be used to
represent particular values of the random variables. Any particular
entry of the vector of probabilities ${ P}_{ABE}$ will thus be
expressed as ${ P}_{ABE}(a,b,e)$. For convenience, probabilities are
allowed to be un-normalized, that is, the only constraint on ${
P}_{ABE}(a,b,e)$ is that all its entries are non-negative. Alice and
Bob are allowed to perform general local operations, where by
general it is meant that the operation need not always be
successful. Alice's operations can be expressed as a $d'_A\times
d_A$ matrix of non-negative entries, denoted by $\D_A(a',a)$, where
$a'\in \{0,...,d'_A-1\}$, $a\in\{0,...,d_A-1\}$
and $\D_A(a',a)\geq 0$. 
 With
probability $\D_A(a',a)$ the output $a$ is written to $a'$. Even
when normalised, the sum of the elements in each column can be less
than one; this expresses the fact that the operation can fail. Bob's
operations are defined by a similar matrix $\G_B$. When $\D_A$ and
$\G_B$ are applied to $P_{ABE}$, the components of the resulting
distribution are denoted by $[\D_A \G_B P_{ABE}](a,b,e)$. In the
event that there is no output after filtering, Alice and Bob
communicate publicly and throw away their data. We now provide
specific definitions of the quantities considered in the rest of the
paper.

\vs {\it Definition 1 [Secret bit fraction of a binary
distribution].} A distribution where $d_A=d_B=2$ and $d_E$ is
arbitrary, is called `binary'. The  secret-bit fraction of the
normalized binary distribution ${  P}_{ABE}$ will be called
$\lambda[{ P}_{ABE}]$. $\lambda[{ P}_{ABE}]$ is the maximum $\tau$
such that there exists a decomposition of ${ P}_{ABE}$ of the form:
\begin{equation}\label{1}
{  P}_{ABE}=\tau {  S}_{AB}{  Q}_E + (1-\tau){  H}_{ABE},
\end{equation}
where $\tau\in[0,1]$. ${  Q}_E$ and ${ H}_{ABE}$ can be any
probability distributions and ${
S}_{AB}(a,b)=\frac{1}{2}\;\delta_{ab}$ is a shared bit. The result
proved in the following lemma will be used widely in this paper.

\vs{\it Lemma 1. \label{lemma1}} Given a binary distribution ${
P}_{ABE}$ (not necessarily normalized) its secret-bit fraction is
the following:
\begin{equation}
    \lambda[{  P}_{ABE}]=
    \frac{2\sum_e \min[{  P}_{ABE}(0,0,e),{  P}_{ABE}(1,1,e)]}
    {\sum_{abe} {  P}_{ABE}(a,b,e)}
\label{sbf}\end{equation}

\vs{\em Proof:} Notice that $\lambda[\nu P]=\lambda[P]$ for any
$\nu>0$. Hence we can assume that $P$ is normalized and forget the
denominator. Taking the optimal decomposition (\ref{1}) and using
the fact that the components of ${ H}_{ABE}$ are positive, one can
write the following componentwise inequality ${ P}_{ABE} \geq
 \tau{  S}_{AB}{ Q}_E$.
Here we have treated ${ P}_{ABE}$ and ${ H}_{ABE}$ as  vectors and
${  S}_{AB}{ Q}_E$ as the tensor product of two vectors. Let ${
Q'}_E\equiv \tau { Q}_E$ (recall $\sum_e{ Q}_E(e) =1$). It follows
that ${ P}_{ABE}(a,b,e)\geq \frac{1}{2} \;\delta_{ab}{ Q}'_E(e)$. If
$a\neq b$ the inequality is satisfied. If $a=b$, then both ${
P}_{ABE}(0,0,e)\geq \frac{1}{2} \;{ Q}'_E(e)$ and ${
P}_{ABE}(1,1,e)\geq \frac{1}{2}\; {  Q}'_E(e)$ must hold. It is
clear that the maximum $\tau$ is achieved with ${ Q}'_E=2 \min[{
P}_{ABE}(0,0,e),{ P}_{ABE}(1,1,e)]$. Substituting this value of ${
Q}'_E(e)$ into $\sum_e{  Q}'_E(e) =\tau$ completes the proof. \hfill
$\blacksquare$

\vs Now, we want to generalize the notion of secret-bit fraction for general distributions, not necessarily being binary. For this we proceed as follows. Given a distribution $P_{ABE}$ (not necessarily binary), we consider all SLOPC protocols whose result is a binary distribution. Among all these binary distributions obtainable from $P_{ABE}$ by SLOPC we want to find the one which maximizes
the formula (\ref{sbf}). Without loss of
generality, any SLOPC protocol can always be decomposed in the
following way. Alice performs the local operation $\D^{(0)}_A$ and
makes public some of her information. One can think that the outcome
of $\D^{(0)}_A$ has two variables $(a',c_1)$, where $a'$ is kept
secretly by Alice, and $c_1$ is broadcasted. Later, Bob, depending
on the message $c_1$ performs a local operation $\G^{(c_1)}_B$ with
outcome $(b',c_2)$, and sends the message $c_2$. Later, Alice,
depending on the messages $c_1 c_2$ performs another local operation
$\D_A^{(c_1 c_2)}$, and so on. If at the end of the protocol none of
Alice's and Bob's operations has failed, for each string of messages
$\bar{c}=(c_1 c_2 c_3 \cdots)$, Alice has performed a string of
operations $\D_A^{(0)} \D_A^{(c_1c_2)} \D_A^{(c_1 c_2 c_3
c_4)}\cdots$. We denote the product of these matrices by
$\D_A^{\bar{c}}$, where the dependence on the public messages is
expressed through $\bar{c}$. Similarly, we define $\G_B^{\bar{c}}$
for Bob. If the initial distribution is $P_{ABE}$, then the final
distribution is $P'_{ABEC}(a,b,e,\bar{c})= \left[ \D_A^{\bar{c}}
\G_B^{\bar{c}} P_{ABE}\right]\!(a,b,e)$ (
here $a$ and $b$ are
binary variables). Having settled all this notation for protocols
with communication, we are ready to prove that communication is not
necessary at all.

\vs {\it Lemma 2.} In order to find the SLOPC protocol that
maximizes $\lambda$, one  need only consider protocols without
public communication.

\vs {\em Proof:} Suppose that at the end of a general SLOPC protocol
the  distribution obtained is $P'_{ABEC}(a,b,e,\bar{c})$, which we
can assume to be normalized. Because the random variable $\bar{c}$
is public, we have to consider it as part of Eve's knowledge
$(e,\bar{c})$. Using formula (\ref{sbf}), the secret-bit fraction of
$P'_{ABEC}(a,b,e,\bar{c})$ satisfies


   \begin{multline}
   \lambda[P'_{ABEC}] =
    2\sum_{e,\bar{c}}\min[P'_{ABEC}(0,0,e,\bar{c}),P'_{ABEC}(1,1,e,\bar{c})]
    \\ =
    \sum_{\bar{c}} P_C(\bar{c})\, \lambda\!\left[P'_{ABEC}(\cdot |\bar{c})\right]
    \\ \leq
    \max_{\bar{c}} \lambda\!\left[P'_{ABEC}(\cdot |\bar{c})\right]\ ,
    \label{max}
    \end{multline}

where $P'_{ABEC}(\cdot |\bar{c})$ denotes the probability
distribution for $ABE$ conditioned on a particular string of
messages $\bar{c}$. If the maximum in (\ref{max}) is attained for
the value $\bar{c}_0$, the protocol without communication consisting
of just the local operations $\D_A^{\bar{c}_0}$ and
$\G_B^{\bar{c}_0}$, is not worse than the general one. \hfill
$\blacksquare$

\vs Lemma 2 allows for a simple mathematical definition of the
principal quantity studied in this paper.

\vs {\it Definition 2 [The  MESBF of a  distribution].} The MESBF of
$P_{ABE}$ is
\begin{equation}\label{mightymice}
    \Lambda[P_{ABE}]= \sup_{\D_A \G_B}
    \lambda[\D_A\G_B P_{ABE}]\ .
\end{equation}


\vs The fact that a supremum, rather than a maximum, is considered
in this definition, follows from the requirement that SLOPC
transformations must succeed with probability strictly larger than
zero. In some cases, the optimal SLOPC transformation does not
exist. But one can apply a transformation giving a secret-bit
fraction as close as one wishes to $\Lambda$. (A very similar
phenomenon appears for the `singlet fraction' of quantum states
\cite{horosinglet} and is called quasi-distillability.)  For any
distribution, $P_{ABE}$, we know that $\Lambda[{ P}_{ABE}]\in
[\frac{1}{2},1]$. The lower bound of $\frac{1}{2}$ can always be
obtained if Alice and Bob throw away any data they have and simply
toss unbiased coins. An important fact about $\Lambda$ is that it is
a secrecy monotone.

\vs {\it Theorem 1.} The quantity $\Lambda[P_{ABE}]$ has the
following properties:
\begin{itemize}
    \item $\Lambda[P_{ABE}]$ is nonincreasing when the honest
    parties perform local operations and public communication.
    Even if these operations can fail with some probability (SLOPC).
    \item $\Lambda[P_{ABE}]$ is nondecreasing when Eve performs
    local operations.
\end{itemize}

\vs {\em Proof:} The proof of the first statement comes from the
 definition of $\Lambda$, in terms of an optimization over all
possible SLOPC protocols. The second statement can be shown by
applying an arbitrary operation $\YY_E$ to Eve's data, and see how
$\lambda$ changes. $\YY_E$ must not be a filtration, because Eve
cannot make the honest parties reject their data.

\begin{multline}\label{}
    \lambda\left[ \YY_E P_{ABE} \right]
    = 2 \sum_{e'} \min[ \sum_e \YY_E(e',e)P_{ABE}(0,0,e),\\
    \sum_e \YY_E(e',e)P_{ABE}(1,1,e)] \\
    \geq 2 \sum_{e',e} \YY_E(e',e) \min\left[ P_{ABE}(0,0,e),
    P_{ABE}(1,1,e) \right] \\
    = 2 \sum_{e} \min\left[ P_{ABE}(0,0,e),
    P_{ABE}(1,1,e) \right]\ .
\end{multline}

Where the inequality comes from the concavity of the $\min$
function. $\blacksquare$

\section{A sufficient condition for distillable secrecy}\label{secIIkey}

In this section we provide a sufficient condition for a
distribution $P_{ABE}$ to allow a strictly positive secret key
rate between Alice and Bob. Performing  collective operations on
sufficient samples from a distribution satisfying this condition,
and by communicating over their insecure channel, Alice and Bob
can always obtain secret key.

\vs{\it Theorem 2.} If $\Lambda[P_{ABE}]>\frac{1}{2}$ then $P_{ABE}$
has distillable secret key.

\vs If filtrations ${ \D}_A$ and $\G_B$ can be found such that
$\lambda[{ \D}_A\G_BP_{ABE}]>\frac{1}{2}$ then $P_{ABE}$ has
distillable key. The proof of this theorem is found in Appendix
\ref{theo2}. There, we describe a protocol with which one can always
distill a secret key, if the condition of the theorem is satisfied.


On completion of this paper we were made aware of the work of
Holenstein \cite{Holenstein05,Holenstein06}. His work defines two
parameters $(\epsilon, \delta)$ associated with each probability distribution $P_{ABE}$
and provides a necessary and sufficient condition for the
distribution to have distillable key in terms of these two
parameters. Given a binary distribution $P_{ABE}$ such that
\begin{eqnarray}
  P_{A}(0) = P_{B}(0) = P_{A}(1) = P_{B}(1) = \frac{1}{2} \\
  P_{AB}(0,0) = P_{AB}(1,1) \geq \frac{1+\epsilon}{2}
\end{eqnarray}
there exists an event $\mathcal{E}$ which implies $A=B$ such that
\begin{eqnarray}
  P_{ABE}(\mathcal{E} | A=B) \geq \delta \\
  I(A:E|\mathcal{E}) =0
\end{eqnarray}
With these definitions we can get the lower bound
\begin{equation}\label{}
    \lambda[P_{ABE}] \geq P_{ABE}(A=B, \mathcal{E} ) \geq \frac{1+\epsilon}{2}\,\delta
\end{equation}
Hence, the distillability
condition in terms of $\Lambda$ follows from Holenstein's condition
in terms of these two parameters.
However, it is insightful to have the distillability condition in
terms of a single quantity, which is an operationally meaningful
secrecy monotone. We should note that $\Lambda[P_{ABE}]$ is defined
through an \emph{optimization} over filtrations unlike Holenstein's
two parameters.

\section{MESBF by reversible operations}
\label{secIIIkey}

\vs In this section we introduce a distinction between operations
that degrade the data, and operations that do not. We say that an
operation $\D$ degrades the data, if once it has been applied to the
data there is no probability that the original data can be
recovered. Then, operations that do not degrade the data are called
{\em reversible}. Mathematically, the operation corresponding to the
matrix $\D$ is reversible if its inverse $\D^{-1}$ has nonnegative
entries. Notice that the fact that the inverse exists, does not mean
that the transformation can be undone with probability one; since
rates of distillation are of no concern in the scenario considered
in this paper, the probabilistic nature of the reversibility is
irrelevant.

\vs Of course classical information can always be copied, and thus,
recovered whatever transformation is applied to it. But, if within a
particular operation data is copied, this has to be represented in
the matrix corresponding to this operation. It is clear that this
kind of operation is always reversible.

\vs{\it Definition 3 [Reversible stochastic transformations].} A stochastic transformation $\D$ is reversible if its inverse $\D^{-1}$ has non-negative entries. This implies that if a given distribution $P$ is processed with $\D$, we can still recover $P$ (with some probability of success) by applying $\D^{-1}$.

\vs As an instance, let us consider transformations on the set of
two-outcome probability distributions. The inverses of $2\times 2$
matrices can be obtained through the following formula
\begin{equation}
\left[%
\begin{array}{cc}
 w & x \\
 y & z \\
\end{array}
\right]^{-1}=\frac{1}{wz-xy}\left[%
\begin{array}{cc}
 z & -x \\
 -y & w \\
\end{array}
\right].
\end{equation}
It is easy to see that $2\times 2$ operations are reversible if,
and only if, they are diagonal or anti-diagonal. This fact will be
used later.

\vs {\it Definition 4 [Equivalent distributions under reversible
operations].} Two probability distributions are called `equivalent'
if there exists a reversible operation which takes one probability
distribution to the other and viceversa. These equivalence classes have the following useful property.

\vs{\it Lemma 3. \label{lemma2}} Within an equivalence class all
distributions have the same MESBF.

\vs{\it Proof:} Suppose that two equivalent distributions, $P_{ABE}$
and $P'_{ABE}$, have different MESBF: $\Lambda[P_{ABE}] <
\Lambda[P'_{ABE}]$ without loss of generality. This gives a
contradiction, because in the protocol that optimizes
$\Lambda[P_{ABE}]$, one can always perform a first step consisting
of going from $P_{ABE}$ to $P'_{ABE}$. $\blacksquare$

\vs In the following we find the MESBF for binary distributions when
Alice and Bob are restricted to performing reversible operations on
their data. For a distribution ${P}_{ABE}$ we call this quantity
$\Lambda_R[{ P}_{ABE}]$.

\vs {\it Definition 5.} The MESBF with reversible operations $\OO_A$
and $\VV_B$ is
\begin{equation} \label{mightymouse}
    \Lambda_R[P_{ABE}]= \sup_{\OO_A \VV_B} \lambda [\OO_A \VV_B P_{ABE}].\
\end{equation}

\vs{\it Theorem 3.} Given a binary distribution $P_{ABE}$ the
maximum value of $\lambda$, after reversible filtrations, is as
follows:
\begin{multline} \Lambda_R[{
P_{ABE}}]=\max\big\{\\\mbox{zmax}_{e'\in\,
\mathbb{Q}}[\;\;\frac{2\sum_e \min[{  P}(0,0,e),\phi_{e'}{
P}(1,1,e)]}{{ P}(0,0)+ \phi_{e'}{ P}(1,1)+ 2\sqrt{ \phi_{e'}{
P}(0,1){ P}(1,0)}}],\\\mbox{zmax}_{e''\in\,
\mathbb{S}}[\frac{2\sum_e \min[{ P}(0,1,e), \psi_{e''}{
P}(1,0,e)]}{{ P}(0,1)+ \psi_{e''}{ P}(1,0)+ 2\sqrt{ \psi_{e''}{
P}(0,0){ P}(1,1)}}]\;\big\},\label{mesbfr}
\end{multline}
where we have suppressed the indices `$A,B, E$' on the right hand
side so that $P={P}_{ABE}$. The formula is further compressed by
writing $ P(a,b)\equiv \sum_e P(a,b,e)$. We define
$\phi_{e'}\equiv\frac{{ P}(0,0,e')}{{ P}(1,1,e')}$ and
$\psi_{e''}\equiv\frac{{ P}(0,1,e'')}{{ P}(1,0,e'')}$. The set
$\mathbb{Q}$ is the set of all $e$ where both $P(0,0,e)\neq 0$ and
$P(1,1,e)\neq 0$. The set $\mathbb{S}$ is the set of all $e$ where
both $P(1,0,e)\neq 0$ and $P(0,1,e)\neq 0$. The operation
$\mbox{zmax}_{e'\in \mathbb{Q}}$ is constructed in the following
way. It returns the maximum value of its argument as $e'$ is varied
over the set $\mathbb{Q}$. If $\mathbb{Q}$ is empty then the
operation is defined as returning `0'. The operation
$\mbox{zmax}_{e''\in \mathbb{S}}$ is defined similarly with regard
to the set $\mathbb{S}$.

\vs{\it Corollary 1.} In the case where Eve is decoupled, $P_{ABE} =
P_{AB} P_E$, this reduces to:
\begin{equation}\label{l0r}
    \Lambda_R[P_{AB}]=
    \left\{\begin{array}{ll}
      &0\; \mbox{if}\; P(0,0)P(1,1)=
      P(0,1)P(1,0)=0 \\ \\&
      \max\!\Big[
      \left(1+\sqrt{\frac{P(0,1)P(1,0)}{P(0,0)P(1,1)}}\right)^{-1},\\&
      \quad\quad\;\;\left(1+ \sqrt{\frac{P(0,0)P(1,1)}{P(0,1)P(1,0)}}\right)^{-1}
      \Big]\,
      \mbox{otherwise}\\
    \end{array}\right\}
\end{equation}

\vs Note that both Theorem 3 and Corollary 1 have lower bounds of
zero. This is in contrast to $\Lambda[P_{ABE}]\in[1/2,1]$ where
the lower bound can always be obtained if Alice and Bob both
perform the irreversible operation of throwing away all data and
tossing unbiased coins. Since irreversible operations are excluded
in the definition of  $\Lambda_R[P_{ABE}]$ it takes a lower bound
of zero.

\vs {\em Proof of Theorem 3.} Let us consider the supremum
(\ref{mightymouse}) with the constraint that $\D_A,\G_B$ are of the
form $\OO_A= \mbox {diag} (\alpha,\beta)$ and $\VV_B= \mbox {diag}
(\gamma,\delta)$ where $\alpha, \beta, \gamma, \delta
>0$.

{\small\begin{multline}\Lambda_R[P] =\sup_{\OO_A \VV_B}
\frac{2\sum_e \min[{  \alpha\gamma P}(0,0,e),\beta\delta{
P}(1,1,e)]}{{ \alpha\gamma P}(0,0)+ \beta\gamma{ P}(1,0)+
\alpha\delta{ P}(0,1)+\beta\delta{ P}(1,1)}\\= \sup_{r
q}\frac{2\sum_e \min[{ P}(0,0,e),r{ P}(1,1,e)]}{{ P}(0,0)+ q{
P}(1,0)+ \frac{r}{q}{ P}(0,1)+r{ P}(1,1)}
\label{mesbfr1}\end{multline}}

 where $q\equiv \frac{\beta}{\alpha}$
and $r\equiv \frac{\beta\delta}{\alpha\gamma}$. We now label  the
outputs of Eve so that $\frac{P(0,0,i)}{P(1,1,i)}\leq
\frac{P(0,0,i+1)}{P(1,1,i+1)}$ for all $i\in\{0,..,d_E-1\}$ (if
there is an $i$ such that $P(0,0,i)=P(1,1,i)=0$ this should be left
out of the ordering; if $P(0,0)=P(1,1)=0$ then one can readily check
that $\lambda[\OO_A\VV_BP_{AB}]=0$). We will now consider the
function $\lambda[\OO_A \VV_B P_{AB}]$ for different ranges of $r$.
\begin{enumerate}
    \item  For
$r\in[\frac{P(0,0,g)}{P(1,1,g)},\frac{P(0,0,g+1)}{P(1,1,g+1)})$,
$g\in\{0,..,d_E-2\}$, Eq. (\ref{mesbfr1}) can be written as:
\begin{equation}
\frac{2\sum_{e=0}^g { P}(0,0,e)+2r\sum_{e=g+1}^{d_E-1} {
P}(1,1,e)]}{{ P}(0,0)+ q{ P}(1,0)+ \frac{r}{q}{ P}(0,1)+r{ P}(1,1)}.
\end{equation}
    \item When $r\in[0,\frac{P(0,0,0)}{P(1,1,0)})$
the numerator of Eq. (\ref{mesbfr1}) becomes $2r\;P(1,1)$.
    \item When $r\in[\frac{P(0,0,d_E-1)}{P(1,1,d_E-1)},\infty)$ the numerator of Eq. (\ref{mesbfr1}) becomes
    $2\;P(0,0)$.
\end{enumerate}
For each range $1.-3.$ by differentiating with respect to $r$,
holding $q$ constant, one can deduce that the maxima are always at
one of the limits of the specified range of $r$. More precisely, the
global maximum of the function in Eq. (\ref{mesbfr1}) occurs when
$r=\frac{{ P}(0,0,e')}{{ P}(1,1,e')}= \phi_{e'}$ for a particular
$e'\in \{0,...,d_E-1\}$. The $r=0$ and $r=\infty$ limits correspond
to minima.

Restricting the function to the points $r=\phi_{e'}$ one can
differentiate with respect to $q$. Using this one finds that the
maxima occur when $q= \sqrt{\phi_{e'}\frac{P(0,1)}{P(1,0)}}$.
Substituting this into Eq. (\ref{mesbfr1}) one obtains the first
term in the `max' in Eq. (\ref{mesbfr}). The
`zmax$_{e'\in\mathbb{Q}}$' indicates that we vary over all
$e'\in\mathbb{Q}$. Since we know that the $r=0$ and $r=\infty$
limits correspond to minima, $\mathbb{Q}$ is constructed to
exclude these situations from the allowed values of $e'$.

We have found the optimal value of $\lambda[\OO_A \VV_B P]$ given
that $\OO_A$ and  $\VV_B$ are diagonal. This is not yet
$\Lambda_R[{P}]$ since there are other possible reversible
filtrations $\OO_A$ and $\VV_B$.

 In this binary case filtering operations are $2\times 2$ matrices. As noted above such filtrations are reversible   only if they are diagonal or
anti-diagonal matrices. Some thought shows that, by considering the
case $\OO_A$ anti-diagonal and $\VV_B$ diagonal, we will have looked
at all distinct reversible operations.

The case where $\OO_A= \mbox {antidiag} (\alpha,\beta)$ and $\VV_B=
\mbox {diag} (\gamma,\delta)$ can be treated using the tools used in
the case where both matrices were diagonal. One obtains as a result
the other term in the `max' in Eq. (\ref{mesbfr}). Again, the
`zmax$_{e'\in\mathbb{S}}$' indicates that we vary over all
$e'\in\mathbb{S}$. $\blacksquare$

\vs By definition $\Lambda_R[P] \leq \Lambda[P]$ holds in general. A
reasonable question to pose is, for which distributions $P$ is the
inequality  saturated such that $\Lambda_R[P] = \Lambda[P]$? In such
cases, locally degrading the data would not help. In the next
section a class of such distributions is given.

\section{The MESBF from private
correlations}\label{secVkey}

In this section we consider the MESBF when Alice and Bob can have
alphabets of any size but they are uncorrelated with the
eavesdropper. Though its proof is nontrivial, the result contained
in Theorem 4 is intuitive. The optimal protocol is to filter only
two outcomes. The result shows that, except for unusual
distributions described below, filtering operations which introduce
local randomness serve no advantage. This is in contrast with the
result of the next section where we find a role for local
randomization. In addition we find that filtering operations which
take several outcomes to just one (eg. `$4$' $\rightarrow$ `$0$' and
`$5$' $\rightarrow$ `$0$') cannot help.

\vs {\it Theorem 4.} For distributions $P_{AB}$ where Eve is
decoupled, the MESBF is the following:
\begin{equation}\label{l0y}
    \Lambda[P_{AB}]=\max_{a_0,b_0,a_1,b_1}
    \left\{\begin{array}{ll}
      &\frac{1}{2}\;\; \mbox{if}\quad P(a_0,b_0)P(a_1,b_1)=\\&
      P(a_0,b_1)P(a_1,b_0)=0 \\ \\
      &\frac{1}{1+
      \sqrt{\frac{P(a_0,b_1)P(a_1,b_0)}{P(a_0,b_0)P(a_1,b_1)}}}
      \;\;\mbox{otherwise} \\
    \end{array}\right\}
\end{equation}

Where, in the maximization $a_0,a_1\in\{0,1,...,d_A-1\}$ and
$b_0,b_1\in\{0,1,...,d_B-1\}$.

\vs The proof of this Theorem is long and is contained in Appendix
\ref{secretkeyprfthrm3}. In the situation $P(a_0,b_0)P(a_1,b_1)=
      P(a_0,b_1)P(a_1,b_0)=0$ local randomness is useful. Throwing
      away all data and using local, unbiased, coin tosses can
      always
      obtain a secret-bit fraction of $\frac{1}{2}$.

\vs {\it Corollary 2.} For $N$ copies of the distribution $P_{AB}$
(represented as $P_{AB}^N$) where Eve is decoupled the MESBF is:
\begin{equation}\label{l0x}
    \Lambda[P_{AB}^N]=\max_{a_0,b_0,a_1,b_1}
    \left\{\begin{array}{ll}
      &\frac{1}{2}\;\; \mbox{if}\; P(a_0,b_0)P(a_1,b_1)=\\&
      P(a_0,b_1)P(a_1,b_0)=0 \\ \\
      &\frac{1}{1+
      \big(\frac{P(a_0,b_1)P(a_1,b_0)}{P(a_0,b_0)P(a_1,b_1)}\big )^{N/2}}
      \,\mbox{otherwise} \\
    \end{array}\right\}
\end{equation}

\vs {\em Proof:} We first note that the expression for
$\Lambda[P_{AB}]$ in Theorem 3 depends monotonically  on the
quantity $\S=\frac{P(a_0,b_1)P(a_1,b_0)}{P(a_0,b_0)P(a_1,b_1)}$.
When the expression is at a maximum, $\S$ is at a minimum. It is
$\S$ that we will consider in the following. We say that a single
copy of a distribution will have output alphabets of sizes $d_A$ and
$d_B$. For $N$ copies of $P_{AB}$ (the distribution $P^N_{AB}$) $\S$
becomes:
\begin{equation} \label{SQuant}
\S=\frac{P^N (\underbar{a}_0,\underbar{b}_1)P^N
(\underbar{a}_1,\underbar{b}_0)}{P^N
(\underbar{a}_0,\underbar{b}_0)P^N (\underbar{a}_1,\underbar{b}_1)},
\ee where $\underbar{a}$ and $\underbar{b}$ can be viewed as $N$
component vectors with each entry $a^{(i)}$ and $b^{(i)}$ chosen
from alphabets of sizes $d_A$ and $d_B$ respectively. Thus, by
definition $P^N (\underbar{a}_0,\underbar{b}_1)= P^{(1)}
({a}^{(1)}_0,{b}^{(1)}_1)P^{(2)} ({a}^{(2)}_0,{b}^{(2)}_1)...P^{(N)}
({a}^{(N)}_0,{b}^{(N)}_1)$. Where $P^{(i)}=P$ is the original single
copy distribution; the superindex $(i)$  appears for counting
purposes.

Performing a similar decomposition for the other three terms in Eq.
(\ref{SQuant}) and with some rearranging one obtains:
\begin{eqnarray} \label{PEXpand} \S=&&\Big[\frac{P^{(1)}
({a}^{(1)}_0,{b}^{(1)}_1)P^{(1)} ({a}^{(1)}_1,{b}^{(1)}_0)}{P^{(1)}
({a}^{(1)}_0,{b}^{(1)}_0)P^{(1)}
({a}^{(1)}_1,{b}^{(1)}_1)}\Big]\nonumber\\&\times
&\Big[\frac{P^{(2)} ({a}^{(2)}_0,{b}^{(2)}_1)P^{(2)}
({a}^{(2)}_1,{b}^{(2)}_0)}{P^{(2)} ({a}^{(2)}_0,{b}^{(2)}_0)P^{(2)}
({a}^{(2)}_1,{b}^{(2)}_1)}\Big]\times...\nonumber\\&\times
&\Big[\frac{P^{(N)} ({a}^{(N)}_0,{b}^{(N)}_1)P^{(N)}
({a}^{(N)}_1,{b}^{(N)}_0)}{P^{(N)} ({a}^{(N)}_0,{b}^{(N)}_0)P^{(N)}
({a}^{(N)}_1,{b}^{(N)}_1)}\Big].
\end{eqnarray}

The maximum value of $\lambda$ corresponds to the situation where
$\S$ is a minimum. We note that each square-bracketed term in Eq.
(\ref{PEXpand}) is labeled by the superindex $(i)$ and depends on a
different set of outcomes $a^{(i)}_0,a^{(i)}_1,b^{(i)}_0,b^{(i)}_1$.
One can thus minimize each square bracketed term in Eq.
(\ref{PEXpand}) independently. Since all of the probability
distributions labeled $(i)$ are the same, one knows that the optimal
choice of $a^{(1)}_0,a^{(1)}_1,b^{(1)}_0,b^{(1)}_1$ for term $(1)$
will also be the optimum for all terms. Eq. (\ref{PEXpand}) thus
becomes $\S=\Big[\frac{P^{(1)} ({a}^{(1)}_0,{b}^{(1)}_1)P^{(1)}
({a}^{(1)}_1,{b}^{(1)}_0)}{P^{(1)} ({a}^{(1)}_0,{b}^{(1)}_0)P^{(1)}
({a}^{(1)}_1,{b}^{(1)}_1)}\Big]^N$. Dropping the label $(1)$ one
obtains Corollary 2. $\blacksquare$

From Corollary 2 one sees that as $N$ increases $\Lambda[P^N_{AB}]$
converges exponentially to 1 if $P_{AB}$ has distillable secrecy.

\section{The MESBF for general
correlations} \label{sectionket}\label{secVIkey}

We have no formula for the MESBF for general distributions
$P_{ABE}$. In the following section we investigate this  case and
identify a distribution, $P_{ABE}$, where irreversible operations
obtain a higher secret-bit fraction than the value obtained by
reversible ones alone.

\vs Theorem 3 shows that local randomization has virtually no role
in the protocols that maximize the secret-bit fraction when Eve is
decoupled. One might therefore hope that, on introducing Eve, local
randomization remains unnecessary. At first glance, local
randomization in one-shot protocols seems to serve no role in
maximizing the secret-bit fraction.  If Alice and Bob locally
degrade their data one might argue that their secret-bit fraction
would inevitably fall. This is incorrect; in the following we
provide an example in which, if Alice and Bob {\em both} locally
degrade their data, the value of their secret-bit fraction is higher
than if they perform only reversible operations. In general,
reversible operations are not optimal filtrations. As soon as Eve is
introduced, there is thus a larger role for local randomness in
maximizing the secret-bit fraction of a distribution. A motivation
for this result is the following: though Alice and Bob do indeed
become less correlated as a result of local randomization, Eve
becomes {\em even} less correlated than them. Note that local
randomization certainly does have established uses in obtaining good
secret key rates in the multi-copy case \cite{CK}; where local
randomization by {\em one} party can improve the rate.

\vs We will now provide an example where, if Alice and Bob randomize
locally, they can improve their secret-bit fraction over the value
obtained by  optimal reversible filtrations.  Before giving the
example we introduce the following notation. Since distributions on
three variables do not lend themselves to easy graphical
representation, we let
$P_{ABE}=\sum_{abe}P_{ABE}(a,b,e)\,\mathbf{d}_{abe}$ where the
orthonormal vectors $\,\mathbf{d}_{abe}$ $\forall\;\; a,b,e$ consist
of the standard basis. (the vectors each represent deterministic
probability distributions on the variables, where only the outcomes
$a,b,e$ can occur). 
Consider the distribution:

\begin{equation}\label{lemur}
P_{ABE}= \frac{1}{24}\big[(6 \,\mathbf{d}_{000} +6
\,\mathbf{d}_{110}) +
(5\,\mathbf{d}_{011}+5\,\mathbf{d}_{101}+2\,\mathbf{d}_{111})\big].
\end{equation}

Note that in the first round bracketed term Eve has `$0$' and in the
second `$1$'. Applying formula (\ref{mesbfr}) to this distribution,
one obtains $\Lambda_R[P_{ABE}]=\frac{1}{2}$. Actually, if Alice and
Bob do nothing, they already have $\lambda[P_{ABE}]=\frac{1}{2}$ (by
Eq. (\ref{sbf})). If both parties perform the filtration
\begin{equation}\label{}
    \D_A=\G_B=
\left[
\begin{array}{cc}
 1 & \epsilon \\
 0 & 1 \\
\end{array}
\right]
\end{equation}
with $\epsilon \approx 0.01$, the transformed distribution,
$P'_{ABE}$, has $\lambda[P'_{ABE}]>\frac{1}{2}$. In this case the
MESBF is not obtained by reversible operations. Here the
randomization can be viewed as having the effect that it creates a
secret bit between Alice and Bob when Eve has the outcome `$1$'.
That more general irreversible filtrations are required to obtain
the highest secret-bit fraction means that the analytical task of
finding $\Lambda[P_{ABE}]$ is difficult in general. Finding
$\Lambda[P_{ABE}]$ numerically for a given distribution, $P_{ABE}$,
is also difficult as the function to be optimized is not concave.

\section{Conclusion}

In this section we review the results obtained, outline open
questions and provide alternative interpretations of the MESBF.

In this paper we have functionally defined a new measure
$\Lambda[P_{ABE}]$ called the MESBF of $P_{ABE}$ and we showed that
it is a secrecy monotone. We showed that if
$\Lambda[P_{ABE}]>\frac{1}{2}$ then the distribution can be used to
distill secret key. We gave a comprehensive characterization of
$\Lambda[P_{AB}]$ when Eve is decoupled and also in the case of
reversible operations on binary distributions. Using the results for
reversible operations we were able to show that there exist
distributions for which the optimal filtration requires local
degradation of data. 
An open problem is to show that $\Lambda[P_{ABE}]>\frac{1}{2}$ is
not a necessary condition for distillability; if it were necessary
then the MESBF would be a very useful tool for the investigation of
bound information \cite{Renner, Renner03}.


In this paper $\Lambda[P_{ABE}]$ has been treated as a measure to
give us yes/no information about whether $P_{ABE}$ can be used to
distill secret key. It can, however be viewed in two other ways:

\begin{itemize}
    \item There is a restricted communication scenario in which
    filtrations of $P_{ABE}$ which maximize the secret-bit fraction are exactly what the co-operating
    players would like to do in order to make their communication as secret as possible: if the parties attempt a form of
    (a) `running' key generation given (b) unlimited streams of
    source data but (c) finite memories.

    (a) By `running'  we mean that as soon as a successful
    filtration has occurred the random bits are used for
    encryption purposes; they are not stored up and then subject
    to information reconciliation and privacy amplification
    \cite{CK,BBCM,Impagliazzo}. This is, of course, a substantial constraint.

    (b) If there is plenty of source data, the fact that heavy
    filtration might be required to maximize the secret-bit fraction is not a problem.

    (c) Their memories must be finite since we consider optimal
    single shot operations.

    In this applied context, the role of local randomization is
    surprising; if Alice and Bob degrade their data they can
    nonetheless
    improve the secrecy of their communication.
    \item Advantage distillation is a standard first step for
    obtaining secret key from samples from a general distribution $P_{ABE}$.
The single shot
    filtrations that are described here can be viewed as a
    generalization of advantage distillation. A filtration that
    maximizes the secret-bit fraction of a distribution  can be viewed
    as an optimal distillation step (in the scenario where the supply of data is not
    limiting). Note that though the approach acts on only one copy of a distribution this single copy can be viewed as many copies of a lower dimensional distribution. The fact that introducing local randomness can be
    helpful in maximizing the secret-bit fraction raises the
    intriguing possibility that degrading data serves a role in
    generalized advantage distillation. In the example given, both Alice and Bob
    symmetrically add noise. This is distinct from the case
    considered in \cite{CK} where only one party adds noise. A
    future area of research would be to attempt to identify a distribution
    where {\em optimal} filtrations require {\em both} parties to
    degrade their data.

\end{itemize}

Acknowledgements: NJ thanks the EPSRC, BBSRC and Royal Commission
for the Exhibition of 1851, LM thanks EU Project QAP (IST-3-015848)

Biographies: NJ Obtained a PhD in Bristol in 2005 and is a group leader in the Oxford Centre for Integrative Systems Biology, Oxford Physics. LM Obtained the PhD degree in Barcelona at 2004. At present is a postdoc at DAMTP Cambridge University. His research is focused on quantum information theory.

\bibliographystyle{IEEEtran}

\begin{thebibliography}{31}
\expandafter\ifx\csname
natexlab\endcsname\relax\def\natexlab#1{#1}\fi
\expandafter\ifx\csname bibnamefont\endcsname\relax
  \def\bibnamefont#1{#1}\fi
\expandafter\ifx\csname bibfnamefont\endcsname\relax
  \def\bibfnamefont#1{#1}\fi
\expandafter\ifx\csname citenamefont\endcsname\relax
  \def\citenamefont#1{#1}\fi
\expandafter\ifx\csname url\endcsname\relax
  \def\url#1{\texttt{#1}}\fi
\expandafter\ifx\csname urlprefix\endcsname\relax\def\urlprefix{URL
}\fi \providecommand{\bibinfo}[2]{#2}
\providecommand{\eprint}[2][]{\url{#2}}

\bibitem{note2} Note that
this is for unconditional, information theoretic, security. If one
is happy to upper bound
 Eve's computational power then other cryptographic schemes can be used e.g. the R.S.A. scheme \cite{RSA}.



\bibitem{Lluisbound}
A. Ac\'\i n, J. I. Cirac and L. Masanes, ``Multipartite Bound
Information Exists and Can Be Activated", \emph{Phys. Rev. Lett.},
vol. 92, pp. 107903, 2004.









%


\bibitem{BBCM}  C. H. Bennett, G. Brassard and J.-M. Robert, ``Privacy
  amplification by public discussion'', \emph{SIAM Journal on Computing},
  Vol. 17, no. 2, 1988 pp. 210--229


\bibitem{BennConc} C.H. Bennett, H.J. Bernstein, S. Popescu and B.
Schumacher, ``Concentrating partial entanglement by local
operations'', \emph{Phys. Rev. A}, vol. 54, pp. 4707--4711, 1996.



\bibitem{CP} D. Collins and S. Popescu, ``Classical analog of entanglement",
\emph{Phys. Rev. A}, vol. 65, pp. 032321-1--032321-11, 2002.

\bibitem{CK}
I.~Csisz\'{a}r and J.~K\"{o}rner, ``Broadcast channels with
confidential
  messages,'' \emph{IEEE Trans. Inform. Theory}, vol.~24, pp.
  339--348, 1978.









\bibitem{GW}
 N. Gisin
and S. Wolf, ``Linking classical and quantum key agreement: is there
bound information?'' in \emph{Proceedings of CRYPTO 2000, Lecture
Notes in Computer Science} vol. 1880 (Springer-Verlag, Berlin,
2000), p. 482.


\bibitem{GisincryptoRMP}  N. Gisin, G. Ribordy, W.
Tittel, and H. Zbinden, Quantum cryptography, \emph{Reviews of
Modern Physics,} vol. 74,  pp. 145--195, 2002.


\bibitem{Renner} N Gisin, R. Renner, and Wolf; ``Linking Classical and Quantum Key Agreement: Is There a Classical Analog to Bound Entanglement?", \emph{Algorithmica}, Springer-Verlag, vol. 34, no. 4, pp. 389--412,  2002.

\bibitem{Holenstein05} T. Holenstein, ``Key Agreement from Weak Bit Agreement'', Proceedings of the 37th ACM
Symposium on Theory of Computing, pp. 664-673, 2005.

\bibitem{Holenstein06} Thomas Holenstein, ``Strengthening Key Agreement using Hard-Core Sets'' PhD thesis, vol. 7 of ETH Series in Information Security and
Cryptography, (Hartung-Gorre Verlag, Konstanz, 2006).


\bibitem{horobound1}
M. Horodecki, P. Horodecki, and R. Horodecki, ``Mixed-state
entanglement and distillation: is there a `bound' entanglement in
nature?'' \emph{Phys. Rev. Lett.,}  vol. 80, pp. 5239--5242, 1998.


\bibitem{horosinglet}
M. Horodecki, P. Horodecki, and R. Horodecki, ``General
teleportation channel, singlet fraction, and quasidistillation,''
\emph{Phys. Rev.  A.} vol. 60,  pp. 1888–-1898, 1999.




\bibitem{Impagliazzo} R. Impagliazzo, L. A. Levin, and M. Luby, ``Pseudo-random
generation from one-way functions'', Proceedings of the 21st ACM
Symposium on Theory of Computing, pp. 12--24, 1989.


\bibitem{maurer}
U.~M. Maurer, ``Secret key agreement by public discussion from
common
  information,'' \emph{IEEE Trans. Inform. Theory}, vol.~39,
  no.~3, pp. 733--742, 1993.




\bibitem{MW1}
U.M. Maurer and S. Wolf, ``Towards characterizing when
information-theoretic key agreement is possible'', \emph{Advances in
Cryptology-ASIACRYPT '96}, Lecture Notes in Computer Science Vol.
1163 (Springer-Verlag, Berlin, 1996) p. 196.




\bibitem{MW}
U. M. Maurer and S. Wolf, ``Unconditional Secure Key Agreement and
the Intrinsic Information", \emph{IEEE Trans. Inform. Theory
Theory}, vol. 45, pp. 499--514, 1999.


\bibitem{note1} Note: If a distribution $P_{ABE}$ is distillable, there
are a sufficiently large number of copies of it, $N$, such that,
when jointly processed, something close to a secret bit can be
obtained, thus $\Lambda[P^N_{ABE}]>\frac{1}{2}$. Complementarily, if
there exists an $N$ such that $\Lambda[P^N_{ABE}]>\frac{1}{2}$,
Theorem 2 warrants that $P_{ABE}$ is distillable.




\bibitem{Renner03} R. Renner and S. Wolf,
``New bounds in secret-key agreement: The gap between formation and
secrecy extraction'',  in \emph {Proceedings of Advances in
Cryptology  EUROCRYPT 2003: International Conference on the Theory
and Applications of Cryptographic Techniques,} Warsaw, Poland, 2003,
Lecture Notes in Computer Science (Springer-Verlag, Berlin, 2003).



\bibitem{RSA} R. Rivest, A. Shamir,  and L. Adleman, ``A method for obtaining
digital signatures and public key cryptosystems'',
\emph{Communications of the ACM}, vol. 21, 120--126,  1978.




\bibitem{Shannon}
C. E. Shannon, ``Communication theory of secrecy systems'', Bell
Sys. Tech. J., vol. 28,  pp. 656--715, 1949.




\bibitem{Wyner}
A. D. Wyner, The wire-tap channel, Bell Sys. Tech. J. vol. 54,
1355--1387, 1975.














\end{thebibliography}



\appendices

\section{Proof of Theorem 2}\label{theo2}

In this section we provide a proof of Theorem 2. To do so, we
explicitly describe the distillation protocol with which one can
distill secret key from all distributions satisfying the condition
of the theorem. This protocol might not be efficient, but it is
enough for our purposes.

\vs{\it Protocol.} The first part of the protocol is similar to
advantage distillation, a procedure introduced in \cite{maurer}.
Alice and Bob take $N$ samples from their distributions,
respectively, $(a_1,a_2,...,a_N)$ and $(b_1,b_2,...,b_N)$. They
perform the following stochastic transformation on their strings:
\begin{eqnarray}
  01010101\cdots &\longrightarrow& \verb"0" \label{s1}\\
  10101010\cdots &\longrightarrow& \verb"1" \label{s2}\\
  \mbox{other} &\longrightarrow& \verb"reject"
\end{eqnarray}
If both succeed they each keep their final ($N^{th}$) bit, denoted
$a'$ and $b'$. They repeat this procedure many times, obtaining a
long string of pairs $(a',b')$. The reason for alternating 0's and
1's in the above sequences is because, even in the case where Alice
and Bob's marginal is biased, the sequences (\ref{s1}) and
(\ref{s2}) are equiprobable.

The second step of the protocol consists of taking long strings of
pairs $(a',b')$ and performing information reconciliation and
privacy amplification, as described by Csisz\'ar and K\"orner in
\cite{CK}. This second step yields a secret key if, and only if,
\begin{equation}\label{}
    H(a'|b') < H(a'|e)\ ,
\end{equation}
where $H(x|y)$ is the Shannon entropy of the random variable $x$
conditioned on $y$ \cite{CK}, and $e$ represents all the information
that Eve has at the end of the first step.

\vs{\it Theorem 2.} If $\Lambda[P_{ABE}]>\frac{1}{2}$ then $P_{ABE}$
has distillable secret key.

\vs {\em Proof:} As in Section \ref{sectionket}, we represent a
distribution as
$P_{ABE}=\sum_{abe}P_{ABE}(a,b,e)\,\mathbf{d}_{abe}$, where
$\,\mathbf{d}_{abe}$ $\forall\;\; a,b,e$ are orthonormal vectors
from the standard basis. 
Consider the distribution
\begin{multline}\label{marmoset}
P_{ABE}= \mu\left(\frac{1}{2} \,\mathbf{d}_{000} +\frac{1}{2}
\,\mathbf{d}_{110}\right) +
(1-\mu)\Big(\eta_{00}\,\mathbf{d}_{001}\\+\eta_{11}\
\mathbf{d}_{112}+\eta_{01}\,\mathbf{d}_{013}+\eta_{10}\,\mathbf{d}_{104}
\Big)\ ,
\end{multline}
where $\mu\in(1/2,1]$, $\sum_{ab}\eta_{ab}=1$ and $\eta_{ab}\geq 0$.
Note that, by   degrading Eve's  data, all distributions $P_{ABE}$
with the same secret-bit fraction $\mu$ and the same marginal for
Alice and Bob (characterized by $\mu$ and
$\eta_{ab}$) can be obtained from (\ref{marmoset}). 
This means that if the distribution (\ref{marmoset}) has distillable
secret key, then any distribution $P'$ with $\lambda[P']=\mu$ will
have distillable secret key.

In the distribution (\ref{marmoset}), with probability $1-\mu$ Eve
knows Alice and Bob's bits perfectly, and with probability $\mu$ she
only knows that they are perfectly correlated. The probability that
Alice and Bob have a different outcome is $\epsilon= (1-\mu)
(\eta_{01}+\eta_{10}) \leq (1-\mu) < 1/2$. In the following we
consider the first step of the protocol described above. In it, the
honest parties accept their data if they have the string (\ref{s1})
or (\ref{s2}). Let $t$ be the probability that Alice obtains the
string (\ref{s1}); this is the same as the probability that she
obtains (\ref{s2}). The chance that Alice and Bob accept the same
string is $2t(1-\epsilon)^N$, and the chance that they accept
opposite strings is $2t\epsilon^N$. Notice that these are the only
two possibilities that pass the filter, hence, the probability that
both parties accept is $2t(\epsilon^N +(1-\epsilon)^N)$. The
probability that Alice and Bob have different strings conditioned on
the fact that they accept is $\epsilon^N/(\epsilon^N +
(1-\epsilon)^N)$. In other words, Bob's uncertainty about Alice's
data is
\begin{multline}\label{}
    H(a'|b')=
    h\!\!\left(\frac{\epsilon^N}{\epsilon^N + (1-\epsilon)^N}\right)
    \\\approx \frac{\epsilon^N}{\epsilon^N + (1-\epsilon)^N}\
    N\, \log_2\!\left(\frac{1-\epsilon}{\epsilon}\right)\ ,
\end{multline}
where $h(r)$ is the Shannon entropy of the distribution $(r, 1-r)$,
and the approximation holds when $N$ is large. Eve's probability of
knowing nothing, conditioned on the fact that Alice and Bob have
publicly accepted a round of the procedure, is
$\mu^N/(\epsilon^N+(1-\epsilon)^N)$. Hence, her uncertainty about
Alice's data is
\begin{equation}\label{}
    H(a'|e)= h\!\!\left(\frac{\mu^N}{\epsilon^N+(1-\epsilon)^N}\right)\ .
\end{equation}
The condition for the functioning of the second step of the
distillation protocol is that Bob's uncertainty $H(a'|b')$ is
strictly smaller than Eve's uncertainty $H(a'|e)$. Due to the fact
that $\epsilon \leq 1-\mu < \mu$ there exists a sufficiently large
$N$ for which $H(a'|b') < H(a'|e)$ holds. $\blacksquare$

\section{Decomposition of general operations \label{decompsec}}

In this section we see how a general operation can be decomposed
into a product of more elementary operations. This decomposition
will be used in the proof of Theorem 4. We will use the notation
from  Section \ref{sectionket}. A matrix $\mathcal M$ can be written
as $ \sum_{ij}\mathcal
M_{ij}\,\mathbf{d}_{i}\!\,\mathbf{d}^{\dagger}_{j}$ where
$\,\mathbf{d}_{i}\!\,\mathbf{d}^{\dagger}_{j}$ is an outer product
between the orthonormal vectors $\,\mathbf{d}_{i}$ and
$\,\mathbf{d}_{j}$ from the standard basis. Note that here the
vectors $\,\mathbf{d}_{i}$ correspond to a deterministic
distribution for just one party (say Alice) and thus only one
subindex is used.

The most general filtering operation with input $\c\in\{1,...,d\}$,
and a bit as output, is
\begin{equation}\label{filter}
    \D=\sum_{\c=0}^{d-1} \big( \D_{0\c} \,\mathbf{d}_{0}+\D_{1\c} \,\mathbf{d}_{1}
    \big)\!\,\mathbf{d}^{\dagger}_{\c},
\end{equation}
with coefficients $\D_{0\c},\D_{1\c}\geq 0$, and $\D_{0\c}+
\D_{1\c}\leq 1$ for all $\c\in\{1,...,d\}$. For each input $\c$, we
specify the bias of its corresponding output with the following
function:
\begin{equation}\label{}
    \omega_{\c}=\left\{
    \begin{array}{ll}
      0 & \mbox{ if}\quad \D_{0\c} \geq \D_{1\c}\\
      1 & \mbox{ if}\quad \D_{0\c} < \D_{1\c} \\
    \end{array}
    \right\}.
\end{equation}For each input $\c$, we quantify how mixed its corresponding
output is with the following quantity:
\begin{equation}\label{mix}
    \mu_{\c}= \left\{
    \begin{array}{ll}
      0 & \mbox{ if}\quad \D_{0\c} = \D_{1\c}=0\\
      1-\frac{\D_{{w_{\c}}\c}}{\D_{0\c}+\D_{1\c}} & \mbox{ otherwise} 
    \end{array}
    \right\}.
\end{equation}
The larger $\mu_{\c}$ is, the more mixed the output (when we input
$\c$). Now, we relabel the input in the following way. First, we
order the values of $\c\in\{1,...,d\}$ with decreasing mixing, that
is, $\mu_{\c}\geq \mu_{\c+1}$ for $\c=0\dots d-1$. Second, we shift
the value of the input by adding 2: $\c\rightarrow \c+2$. Let us
denote a generic mixing matrix by:

\begin{multline}
    M(\mu)=(1-\mu)(\,\mathbf{d}_{0}\!\,\mathbf{d}^{\dagger}_{0}+\,\mathbf{d}_{1}\!\,\mathbf{d}^{\dagger}_{1})+\mu(\,\mathbf{d}_{0}\!\,\mathbf{d}^{\dagger}_{1}+\,\mathbf{d}_{1}\!\,\mathbf{d}^{\dagger}_{0}),\\
    \quad\mbox{with}\quad\mu\in[0,1/2].
\end{multline}

It is clear that we can write Eq. \w{filter} as
\begin{equation}\label{D}
    \D=\sum_{\c=2}^{d+1}\, (\D_{0\c}+\D_{1\c})\, M(\mu_{\c})
    \,\mathbf{d}_{\omega_{\c}}\!\,\mathbf{d}^{\dagger}_{\c},
\end{equation}
where the argument of $M(\mu_{\c})$ is the mixing of input $\c$, Eq.
\w{mix}. Consider a $(d+2)$-dimensional linear space with basis
vectors $\{\,\mathbf{d}_{0},\,\mathbf{d}_{1},\ldots
\,\mathbf{d}_{d},\,\mathbf{d}_{d+1}\}$. The vectors
$\{\,\mathbf{d}_{2},\ldots \,\mathbf{d}_{d},\,\mathbf{d}_{d+1}\}$
correspond to the input, and, the vectors
$\{\,\mathbf{d}_{0},\,\mathbf{d}_{1}\}$ correspond to the output.
The matrix \w{D} can be viewed as a square matrix in this
$(d+2)$-dimensional space, with all the non-zero elements contained
in a $2\!\times\! d$ sub-matrix.

In this larger space we define the square matrices
\begin{eqnarray}
    \R &=&\sum_{\c'=2}^{d+1}\, (\D_{0\c'}+\D_{1\c'})\, \proj{\c'} \\
    \mathcal{G}_{\c}&=&\I +\,\mathbf{d}_{\omega_{\c}}\!\,\mathbf{d}^{\dagger}_{\c} \label{whatisG}\\
    \mathcal{W}_{\c}&=&(1-\nu_{\c})(\,\mathbf{d}_{0}\!\,\mathbf{d}^{\dagger}_{0}+\,\mathbf{d}_{1}\!\,\mathbf{d}^{\dagger}_{1})\nonumber\\&&+\nu_{\c}(\,\mathbf{d}_{0}\!\,\mathbf{d}^{\dagger}_{1}+\,\mathbf{d}_{1}\!\,\mathbf{d}^{\dagger}_{0})+\I_{\{2,...,d+1\}}
\end{eqnarray}
for $\c=2,...,d+1$. The numbers $\nu_{\c}$ lie within the range $[0,
1/2]$. If a matrix has the subindex $\{\c_1,\c_2,\ldots\}$, it is
understood that it only has support on the subspace spanned by
$\{\,\mathbf{d}_{\c_1},\,\mathbf{d}_{\c_2},\ldots\}$. For example,
$\I$ is the identity matrix on the whole space, whilst
$\I_{\{0,1\}}=\proj{0} + \proj{1}$. One can readily  check the
following identity:
\begin{eqnarray}\label{product}
    &&\I_{\{0,1\}}\, \mathcal{W}_{d+1}\, \mathcal{G}_{d+1}\, \cdots \mathcal{W}_2\, \mathcal{G}_2\, \I_{\{2,...,d+1\}}\\&&=\mathcal{W}_{d+1}\,\mathbf{d}_{\omega_{d+1}}\!\,\mathbf{d}^{\dagger}_{d+1}
    +[\mathcal{W}_{d+1}\mathcal{W}_d]\,\mathbf{d}_{\omega_{d}}\!\,\mathbf{d}^{\dagger}_{d}
    \nonumber\\&&+\cdots +[\mathcal{W}_{d+1}\mathcal{W}_d\cdots \mathcal{W}_2]\,\mathbf{d}_{\omega_2}\!\,\mathbf{d}^{\dagger}_{2}
    \nonumber
\end{eqnarray}
We have not yet specified the parameters $\nu_{\c}$. If we set
$\nu_{d+1}=\mu_{d+1}$, then
\begin{equation}\mathcal{W}_{d+1}\,\mathbf{d}_{\omega_{d+1}}\!\,\mathbf{d}^{\dagger}_{d+1}=M(\mu_{d+1})\,\mathbf{d}_{\omega_{d+1}}\!\,\mathbf{d}^{\dagger}_{d+1}.\nonumber\end{equation} By construction, we know
that $\mu_d\geq\mu_{d+1}$. Hence, because the matrices $M(\mu)$
commute, we can assign to $\nu_d$ the value such that
$\mathcal{W}_{d+1}\mathcal{W}_d\,\mathbf{d}_{\omega_{d}}\!\,\mathbf{d}^{\dagger}_{d}=
M(\mu_{d})\,\mathbf{d}_{\omega_{d}}\!\,\mathbf{d}^{\dagger}_{d}$. In
the same fashion, we can obtain the values for all the parameters
$\{\nu_2,\ldots \nu_{d+1}\}$ such that
$[\mathcal{W}_{d+1}\mathcal{W}_d\cdots
\mathcal{W}_{\c}]\,\mathbf{d}_{\omega_{\c}}\!\,\mathbf{d}^{\dagger}_{\c}=M(\mu_{\c})\,\mathbf{d}_{\omega_{\c}}\!\,\mathbf{d}^{\dagger}_{\c}$,
for $\c=2,...,d+1$. Finally, we can write the full decomposition of
Eq.  \w{D}:
\begin{equation}\label{thingortwo}
    \D=\I_{\{0,1\}}\, \mathcal{W}_{d+1}\, \mathcal{G}_{d+1}\, \cdots \mathcal{W}_2\, \mathcal{G}_2\, \R
\end{equation}

In the next section it will prove useful to have a decomposition of
$ M(\mu)$. It is clearer to use conventional matrix notation here.
\begin{eqnarray}
   M(\mu)&&=\left[%
\begin{array}{cc}
 1-\mu & \mu \\
 \mu & 1-\mu \\
\end{array}
\right]\nonumber\\&&=\left[%
\begin{array}{cc}
 1-\mu & 0 \\
 0 & 1-\mu \\
\end{array}
\right]\left[%
\begin{array}{cc}
 1 & \frac{\mu}{1-\mu} \\
 \frac{\mu}{1-\mu} & 1 \\
\end{array}
\right]
\end{eqnarray}
this can be further decomposed by noting that:
\begin{multline}
 \left[%
\begin{array}{cc}
 1 & \frac{\mu}{1-\mu} \\
 \frac{\mu}{1-\mu} & 1 \\
\end{array}
\right]= \left[%
\begin{array}{cc}
 1 & 0 \\
 \frac{\mu}{1-\mu} & 1 \\
\end{array}
\right] \left[%
\begin{array}{cc}
 1 & 0 \\
 0 & 1-(\frac{\mu}{1-\mu})^2 \\
\end{array}
\right]\\\times \left[%
\begin{array}{cc}
 1 & \frac{\mu}{1-\mu} \\
 0 & 1 \\
\end{array}
\right].
\end{multline}
We will also use the fact that:
\begin{equation}
\left[%
\begin{array}{cc}
 1 & \frac{\mu}{1-\mu} \\
 0 & 1 \\
\end{array}
\right]=\left[%
\begin{array}{cc}
 0 & 1 \\
 1 & 0 \\
\end{array}
\right]\left[%
\begin{array}{cc}
 1 & 0 \\
 \frac{\mu}{1-\mu} & 1 \\
\end{array}
\right]\left[%
\begin{array}{cc}
 0 & 1 \\
 1 & 0 \\
\end{array}
\right].
\end{equation}

The operations $\mathcal{W}_c$ can thus be expanded as:
\begin{multline}
\mathcal{W}_c=
\left[%
\begin{array}{cc}
 1-\nu_{c} & 0 \\
 0 & 1-\nu_{c} \\
\end{array}
\right]\left[%
\begin{array}{cc}
 1 & 0 \\
 \frac{\nu_{c}}{1-\nu_{c}} & 1 \\
\end{array}
\right]\\\times\left[%
\begin{array}{cc}
 1 & 0 \\
 0 & 1-(\frac{\nu_{c}}{1-\nu_{c}})^2 \\
\end{array}
\right] \left[%
\begin{array}{cc}
 0 & 1 \\
 1 & 0 \\
\end{array}
\right]\\\times\left[%
\begin{array}{cc}
 1 & 0 \\
 \frac{\nu_{c}}{1-\nu_{c}} & 1 \\
\end{array}
\right]\left[%
\begin{array}{cc}
 0 & 1 \\
 1 & 0 \\
\end{array}
\right]_{\{0,1\}}\\ +\mathcal{I}_{\{2,...,d+1\}}
\end{multline}

Since this decomposition of $\mathcal{W}_c$ will be used repeatedly
in the following proof we will need to express it more compactly as:
\begin{equation}
\mathcal{W}_c=\mathcal{K}^{(1)}_{c}\mathcal{T}_c\mathcal{K}^{(2)}_{c}\mathcal{K}^{(3)}\mathcal{T}_c\mathcal{K}^{(3)}\label{whereswashboard}\ee
where
\begin{eqnarray}
\mathcal{K}^{(1)}_{c}&&=\left[%
\begin{array}{cc}
 1-\nu_{c} & 0 \\
 0 & 1-\nu_{c} \\
\end{array}
\right]_{\{0,1\}}+\mathcal{I}_{\{2,...,d+1\}}\\
\mathcal{K}^{(2)}_{c}&&=\left[%
\begin{array}{cc}
 1 & 0 \\
 0 & 1-(\frac{\nu_{c}}{1-\nu_{c}})^2 \\
\end{array}
\right]_{\{0,1\}}+\mathcal{I}_{\{2,...,d+1\}}\\
\mathcal{K}^{(3)}&&=\left[%
\begin{array}{cc}
 0 & 1 \\
 1 & 0 \\
\end{array}
\right]+\mathcal{I}_{\{2,...,d+1\}}\\
\mathcal{T}_{c}&&=\left[%
\begin{array}{cc}
 1 & 0 \\
 \frac{\nu_{c}}{1-\nu_{c}} & 1 \\
\end{array}
\right]+\mathcal{I}_{\{2,...,d+1\}} \label{MrT}
\end{eqnarray}

\section{Proof Of Theorem 4}\label{secretkeyprfthrm3}
In this section we prove Theorem 4. The decomposition provided in
the previous section will be used extensively. We first define more
useful quantities, then derive some useful consequences and finally
provide the proof.

\subsection{Definitions}
In the previous section we showed that filtrations $\D$, represented
by $2\times d$ matrices, can be expressed as $(d+2)\times (d+2)$
matrices. These were then decomposed into products of square
matrices as in Eq. (\ref{thingortwo}). Analogously, we will express
$P_{AB}$ in this larger space. We construct the $(d+2)\times (d+2)$
matrix $\bar{P}_{AB}$ from $P_{AB}$ as follows:
\begin{equation}\label{l0ew}
   \bar{P}_{AB}(a,b)=
    \left\{\begin{array}{ll}
      &0,  \mbox{ if either } a \mbox{ or } b\in\{0,1\} \\ \\
      &P_{AB}(a-2,b-2), \mbox{ otherwise }\\
    \end{array}\right\}
\end{equation}
  for $a\in\{0,...,d_{A}+1\} \mbox{ and } b
\in\{0,...,d_{B}+1\}$.

We now define a function on general probability distributions,
$P_{AB}$, which have $a\in\{0,...,d_{A}+1\} \mbox{ and } b
\in\{0,...,d_{B}+1\}$. These general distributions need not satisfy
the promise in Eq. (\ref{l0ew}) that ${P}_{AB}(a,b)= 0  \mbox{ if
either } a \mbox{ or } b\in\{0,1\}$.

{\it Definition 6. [The function $\vartheta[P_{AB}]$].} Consider a
probability distribution with entries $P_{AB}(a,b)$, where
$a\in\{0,1,...,d_A+1\}$ and $b\in\{0,1,...,d_B+1\}$. Let us define
the following quantity:
\begin{equation}\label{Icantellyoua}
    \vartheta[{P}_{AB}]=\max_{a_0,b_0,a_1,b_1}
    \left\{\begin{array}{ll}
      &\frac{1}{2}\;\; \mbox{if}\quad P(a_0,b_0)P(a_1,b_1)=\\&
      P(a_0,b_1)P(a_1,b_0)=0 \\ \\
      &\frac{1}{1+
      \sqrt{\frac{P(a_0,b_1)P(a_1,b_0)}{P(a_0,b_0)P(a_1,b_1)}}}
      \;\;\mbox{otherwise} \\
    \end{array}\right\},
\end{equation}
where, in the maximization $a_0,a_1\in\{0,1,...,d_A+1\}$ and
$b_0,b_1\in\{0,1,...,d_B+1\}$.

We remark that if the distribution $P_{AB}$ were not normalized, its
value of $\vartheta$ will be unchanged. $\vartheta$ is thus well
defined on un-normalized or filtered distributions.

 We will also
define a modified form of $\D$:
\begin{equation}\label{shesays}
    \bar{\D}=\D+\mathcal{I}_{\{2,...,d+1\}}.
\end{equation}
Given $\D_A$ and $\G_B$ we can find $\bar{\D}_A$ and $\bar{\G}_B$ as
above. As noted above we can also form $\bar{P}_{AB}$  for
$a\in\{0,...,d_{A}+1\} \mbox{ and } b \in\{0,...,d_{B}+1\}$ from the
 distribution $P_{AB}$ using Eq. (\ref{l0ew}). We now
note that:

\begin{enumerate}
    \item $(\bar{\D}_A\bar{\G}_B\bar{P}_{AB})(a,b)=(\D_A\G_BP_{AB})(a,b)$ for
    $a,b\in\{0,1\}$
    \item
    $(\bar{\D}_A\bar{\G}_B\bar{P}_{AB})(a,b)=\bar{P}_{AB}(a,b)={P}_{AB}(a-2,b-2)$
    for $a\in\{2,...,d_A+1\}$ and $b\in\{2,...,d_B+1\}$.
    \item $(\bar{\D}_A\bar{\G}_B\bar{P}_{AB})(a,b)=0$ otherwise.
\end{enumerate}
Here, an expression of the form
$(\bar{\D}_A\bar{\G}_B\bar{P}_{AB})(a,b)$, identifies the entry
$(a,b)$ of the un-normalized matrix yielded by the filtrations
$\bar{\D}_A\bar{\G}_B$ on $\bar{P}_{AB}$.

\subsection{Preparatory remarks and lemmas} \label{prfthrm3}
In this subsection we will prove a few basic results using the
objects defined in the previous subsection. These will then be
applied in the next subsection to prove Theorem 4.


We will now show that:\begin{equation} \label{myfriendgooo}
\Lambda_R[\D_A\G_BP_{AB}]=\vartheta[\bar{\D}_A \;
\bar{\G}^{\opt}_B\;\bar{P}_{AB}],
\end{equation} where $\bar{\D}^{\opt}_A$ is formed from $\D^{\opt}_A$ as in Eq.
(\ref{shesays}) and $\bar{\G}^{\opt}_B$ similarly. The distribution
$\bar{P}_{AB}$ is formed from ${P}_{AB}$ as in Eq. (\ref{l0ew}). Eq.
(\ref{myfriendgooo}) follows from the fact that
$\bar{\D}^{\opt}_A\;\bar{\G}^{\opt}_B\;\bar{P}_{AB}$ contains the
entries of ${\D}^{\opt}_A\;{\G}^{\opt}_B\;{P}_{AB}$ (as noted in
point $1.$ of the preceding subsection) 
and the fact that Eq. (\ref{Icantellyoua}) is the same function  as
Eq. (\ref{l0r}) if the optimal values of $a_0,a_1,b_0,b_1$ are $0$
and $1$ (Eq. (\ref{l0r}) returns the value of $\Lambda_R[{P}_{AB}]$
if $P_{AB}$ is a binary distribution). 


The following three lemmas will be used in the proof of Theorem 4.

{\it Lemma 4.} When either permutation matrices or   diagonal
matrices with entries in the range $(0,1]$ operate on ${P}_{AB}$,
$\RR_A\VV_B{P}_{AB}$, they leave $\vartheta[ P_{AB}]$ unaltered.
Here $a\in\{0,1,...,d_A+1\}$ and $b\in\{0,1,...,d_B+1\}$ and
$P_{AB}$ is a general distribution on these outcomes.

 {\em Proof:} This can be checked by looking at the structure of the
function $\vartheta$ noting that: (a) since the maximization
condition in $\vartheta$ varies over all $a_0,b_0,a_1,b_1$
permutations on ${P}_{AB}$ have no effect (b) the quantity
$\sqrt{\frac{P(a_0,b_1)P(a_1,b_0)}{P(a_0,b_0)P(a_1,b_1)}}$ is
unaltered by the operations defined by diagonal matrices.
$\blacksquare$

We will introduce the  following definition which will be used in
Lemma 5.
\begin{multline}
\mathcal{T}(r)\equiv\left[%
\begin{array}{cc}
 1 & 0 \\
 r & 1 \\
\end{array}
\right]\\\quad \quad
\quad+\I_{\{2,...,d+1\}}=(\,\mathbf{d}_{0}\!\,\mathbf{d}^{\dagger}_{0}+\,\mathbf{d}_{1}\!\,\mathbf{d}^{\dagger}_{1}))+r\,\mathbf{d}_{1}\!\,\mathbf{d}^{\dagger}_{0}+\I_{\{2,...,d+1\}},
\end{multline}
for $r>0$. Though we call this a `filtration', note that
$\mathcal{T}_{00}+\mathcal{T}_{10} \geq 1$. This relaxed definition
of a filtration will not prove problematic  (one can always
normalize such filtrations if necessary). Note that from Eq.
(\ref{MrT}) $\mathcal{T}_c=\mathcal{T}(\frac{\nu_{c}}{1-\nu_{c}})$.

{\em Lemma 5.} Filtering operations  $\mathcal{T}_A
\mathcal{\mathcal{I}}_B$ on $P_{AB}$ cannot increase
$\vartheta[P_{AB}]$.

{\em Proof:} We first note, as in the Proof to Corollary 2, that
$\vartheta[P_{AB}]$ is a variation over
$\S=\frac{P(a_0,b_1)P(a_1,b_0)}{P(a_0,b_0)P(a_1,b_1)}$ for all
$a_0,a_1\in\{0,1,...,d_A+1\}$ and $b_0,b_1\in\{0,1,...,d_B+1\}$ and
it picks out the minimum $\S$. When $\vartheta[P_{AB}]$ is at a
maximum, $\S$ is at a minimum. It is $\S$ that we will consider in
the following.

For a given distribution, $P_{AB}$, $\S$  takes a minimum for a
particular set of  values
$(a_0=a^{\mbox{o}}_0,a_1=a^{\mbox{o}}_1,b_0=b^{\mbox{o}}_0,b_1=b^{\mbox{o}}_1)$.
Two cases can occur with regards to
$(a^{\mbox{o}}_0,a^{\mbox{o}}_1,b^{\mbox{o}}_0,b^{\mbox{o}}_1)$:

\begin{enumerate}
    \item $a^{\mbox{o}}_0=0$ and, or $a^{\mbox{o}}_1=0$
    \item $a^{\mbox{o}}_0\neq 0$ and $a^{\mbox{o}}_1\neq 0$
\end{enumerate}

Suppose, in Case 1., $a^{\mbox{o}}_0=0$. After the filtering
$\mathcal{T}_A \mathcal{I}_B$, $\S$ becomes:
\begin{equation}
\S(r)=\frac{\big(P(0,b^{\mbox{o}}_1)+r
P(1,b^{\mbox{o}}_1)\big)P(a^{\mbox{o}}_1,b^{\mbox{o}}_0)}{\big(P(0,b^{\mbox{o}}_0)+rP(1,b^{\mbox{o}}_0)\big)P(a^{\mbox{o}}_1,b^{\mbox{o}}_1)}
\ee Since we know that the particular set of values
$(a^{\mbox{o}}_0=0,a^{\mbox{o}}_1,b^{\mbox{o}}_0,b^{\mbox{o}}_1)$
are such as to minimize $\S$, we know that $\S(r=0)\leq
\S(r=\infty)$. It follows, noting how $\S(r)$ depends on $r$, that
$\S(r=0)\leq \S(r)$. In this case $\mathcal{T}_A \mathcal{I}_B$ on
$P_{AB}$ does not decrease $\S$.

Though applying $\mathcal{T}_A \mathcal{I}_B$ can only raise the
$\S$ corresponding to the outputs
$(a^{\mbox{o}}_0,a^{\mbox{o}}_1,b^{\mbox{o}}_0,b^{\mbox{o}}_1)$, it
might be the case that this operation might lower the $\S$ value of
other output sets. In fact, the argument provided above is generic.
It can be used to show that $\mathcal{T}_A \mathcal{I}_B$
filtrations cannot yield an $\S$ value lower than the minimum before
the filtration.

It follows that $\vartheta[P_{AB}]\geq \vartheta[\mathcal{T}_A
\mathcal{I}_BP_{AB}]$.

Similar arguments can be used when $a^{\mbox{o}}_1=0$ or indeed
$a^{\mbox{o}}_0=a^{\mbox{o}}_1=0$.

Case 2 is simpler. The transformation $\mathcal{T}_A \mathcal{I}$
leaves $(a_0,a_1,b_0,b_1)$, and the corresponding $\S$, unaltered
(recall that $\S$ is still valid for unnormalized distributions). In
this case $\vartheta[P_{AB}]= \vartheta[\mathcal{T}_A
\mathcal{I}_BP_{AB}]$. Though other entries of the distribution
$P_{AB}$ will be changed by the filtration, arguments with the same
flavor as those used for Case 1. show that these changes leave
$\vartheta[P_{AB}]$ unaltered. $\blacksquare$

It follows by symmetry that identical statements hold for
filtrations of the form $\mathcal{I}_A\mathcal{T}_B$.

We will now make a definition which will be used in the following
Lemma.\begin{equation}
\mathcal{G'}=\I +r\,\mathbf{d}_{0}\!\,\mathbf{d}^{\dagger}_{\c} .\label{whatisG'}\\
\ee Note that $\mathcal{G'}$ is very close to $ \mathcal{G}_{\c}$ as
defined in Eq. (\ref{whatisG}).

 {\it Lemma 6.} Filtering operations of the form $\mathcal{G'}_A
\mathcal{I}_B$ on $P_{AB}$ cannot increase $\vartheta[P_{AB}]$.

{\em Proof:} This proof is very similar to the proof for the
preceding Lemma. We consider the quantity $\S$ again. There will be
an optimal set of outputs
$(a^{\mbox{o}}_0,a^{\mbox{o}}_1,b^{\mbox{o}}_0,b^{\mbox{o}}_1)$ for
which $\S$ takes a minimum. This time the two cases that need to be
considered are:
\begin{enumerate}
    \item $a^{\mbox{o}}_0=c$ and, or $a^{\mbox{o}}_1=c$
    \item $a^{\mbox{o}}_0\neq c$ and $a^{\mbox{o}}_1\neq c$
\end{enumerate}
In Case 1. if $a^{\mbox{o}}_0=c$. After the filtering
$\mathcal{G'}_A \mathcal{I}_B$, $\S$ becomes:
\begin{equation}
\S(r)=\frac{\big ( P(c,b^{\mbox{o}}_1)+
rP(0,b^{\mbox{o}}_1)\big)P(a^{\mbox{o}}_1,b^{\mbox{o}}_0)}{\big(
P(c,b^{\mbox{o}}_0)+r
P(0,b^{\mbox{o}}_0)\big)P(a^{\mbox{o}}_1,b^{\mbox{o}}_1)} \ee Now,
as in Lemma 4, one uses the fact that $\S(r=0)\leq \S(r=\infty)$ to
show  that $\S(r=0)\leq \S(r)$. The rest of this proof follows along
the same lines as the proof for Lemma 5. $\blacksquare$

\subsection{Proof of Theorem 4} In this section we will prove that
$\Lambda[P_{AB}]=\vartheta[ P_{AB}]$. It is straightforward to see
that, for all $\D_A$ and $\G_B$, $\lambda[\D_A\G_BP_{AB}]\leq
\Lambda_R[\D_A\G_BP_{AB}]$. From the last section we note that
$\Lambda_R[\D_A\G_BP_{AB}]=\vartheta[\bar{\D}_A \;
\bar{\G}^{\opt}_B\;\bar{P}_{AB}]$. In this section we prove that
$\vartheta[\bar{\D}_A \; \bar{\G}^{\opt}_B\;\bar{P}_{AB}]\leq
\vartheta[\bar{P}_{AB}]=\vartheta[{P}_{AB}]$. It follows that
$\lambda[\D_A\G_BP_{AB}]\leq \vartheta[{P}_{AB}]$ for all $\D_A$
and $\G_B$, which implies that $\Lambda[P_{AB}] \leq
\vartheta[P_{AB}]$. On the other hand, the function
$\vartheta[P_{AB}]$ is the secret bit fraction obtained with a
particular (reversible) processing of $P_{AB}$, therefore
$\vartheta[P_{AB}] \leq \Lambda[P_{AB}]$. The previous two
inequalities imply $\Lambda[P_{AB}] = \vartheta[P_{AB}]$, which is
the statement of Theorem 4.



\medskip

The  approach uses the decomposition found in Section
\ref{decompsec} combined with the preceding lemmas to show that all
filtrations will either lower $\vartheta[\bar P_{AB}]$ or leave it
the same. Filtrations $\mathcal{\bar D}_A{\bar \G}_B$ will be
expressed as products of operations \begin{equation}\mathcal{
Q}^{(a)}_A\I_B\,\mathcal{ Q}^{(b)}_A\I_B\,\mathcal{
Q}^{(c)}_A\I_B...\mathcal{ Q}^{(M)}_A\I_B\I_A\mathcal{\bar
D}_B\nonumber.\end{equation} We then show that
\begin{multline}\vartheta[\mathcal{ Q}^{(a)}_A\I_B\,\mathcal{
Q}^{(b)}_A\I_B\,\mathcal{ Q}^{(c)}_A\I_B...\mathcal{
Q}^{(M)}_A\I_B\I_A{\bar \G}_B\bar P_{AB}]\\\leq \vartheta[\mathcal{
Q}^{(b)}_A\I_B\,\mathcal{ Q}^{(c)}_A\I_B...\mathcal{
Q}^{(M)}_A\I_B\I_A{\bar \G}_B\bar P_{AB}]\\ \leq \vartheta[\mathcal{
Q}^{(c)}_A\I_B...\mathcal{ Q}^{(M)}_A\I_B\I_A{\bar \G}_B\bar P_{AB}]
\leq ...\leq \vartheta[\I_A{\bar \G}_B\bar
P_{AB}].\nonumber\end{multline} Similar arguments can then be used
to show $\vartheta[\I_A{\bar \G}_B\bar P_{AB}]\leq\vartheta
[\I_A\I_B\bar P_{AB}]$

\emph{Proof:} The following shows that $\vartheta[\bar{\D}_A \;
\bar{\G}^{\opt}_B\;\bar{P}_{AB}]\leq \vartheta[\bar{P}_{AB}]$.
Consider the filtration operations $\mathcal{D}_A$, $\G_B$.
Each $\mathcal{\bar D}$ can be decomposed according to Eq.
(\ref{thingortwo}). We note, using Eq. (\ref{thingortwo}) to expand
$\mathcal{\bar D}_A$, that:
\begin{multline}\label{biffy}
\mathcal{\bar D}_A{\bar \G}_B\bar P_{AB}=
{\mathcal{W}_{{d_A}+1}}_AI_B\,
{\mathcal{G}_{{d_A}+1}}_AI_B\,{\mathcal{W}_{{d_A}}}_AI_B\\ \cdots
{\mathcal{G}_2}_AI_B\, \R_A I_B\;\;\;I_A{\bar \G}_B
P_{AB}\end{multline}

Each of the $\mathcal{W}_c$ can be decomposed further using Eq.
(\ref{whereswashboard}). 
 Eqs. (\ref{number1}-\ref{number4})
 are successive re-writings of Eq. (\ref{biffy}) which will prove
 useful.\begin{eqnarray}
\mathcal{\bar D}_A{\bar \G}_B\bar P_{AB}&&=
{\mathcal{W}_{{d_A}+1}}_AI_B\,
 {\mathcal{G}_{{d_A}+1}}_AI_BP^{'}_{AB}\label{number1}\\
&&={\mathcal{W}_{{d_A}+1}}_AI_BP''_{AB}\nonumber\\&&={\mathcal{K}^{(1)}_{d+1}}_AI_B{\mathcal{T}_{d+1}}_AI_B{\mathcal{K}^{(2)}_{d+1}}_AI_B\nonumber\\&&\times{\mathcal{K}^{(3)}}_AI_B{\mathcal{T}_{d+1}}_AI_B{\mathcal{K}^{(3)}}_AI_B
P''_{AB}
\\&&=
{\mathcal{K}^{(1)}_{d+1}}_AI_B{\mathcal{T}_{d+1}}_AI_BP'''_{AB}\\&&={\mathcal{K}^{(1)}_{d+1}}_AI_BP''''_{AB}\label{number4}
 \end{eqnarray}
where $P^{'}_{AB}$ $=$ ${\mathcal{W}_{{d_A}}}_AI_B $ $\cdots$ $
{\mathcal{G}_2}_AI_B\, \R_A I_B\,I_A{\bar \G}_B\bar P_{AB}$,
 $P''_{AB}$ $=$ ${\mathcal{G}_{{d_A}+1}}_AI_BP^{'}_{AB}$, $P'''_{AB}$ $=$ ${\mathcal{K}^{(2)}_{d+1}}_AI_B{\mathcal{K}^{(3)}}_AI_B$ ${\mathcal{T}_{d+1}}_AI_B{\mathcal{K}^{(3)}}_AI_B
P''_{AB}$ and finally $P''''_{AB}$ $=$
${\mathcal{T}_{d+1}}_AI_BP'''_{AB}$.


\medskip

The operation ${\mathcal{K}^{(1)}_{d+1}}_A$ is reversible. It
follows, using Lemma 4 and Eq. (\ref{number4}), that
$\vartheta[\mathcal{\bar D}_A{\bar \G}_B\bar
P_{AB}]=\vartheta[{\mathcal{K}^{(1)}_{d+1}}_AI_BP''''_{AB}]=\vartheta[P''''_{AB}]$.

\medskip

We know that
$\vartheta[P''''_{AB}]=\vartheta[{\mathcal{T}_{d+1}}_AI_BP'''_{AB}]$
(since $P''''_{AB}={\mathcal{T}_{d+1}}_AI_BP'''_{AB}$). Now, by
using Lemma 5, it follows that
$\vartheta[{\mathcal{T}_{d+1}}_AI_BP'''_{AB}]\leq
\vartheta[P'''_{AB}]$. It follows that $\vartheta[\mathcal{\bar
D}_A{\bar \G}_B\bar
P_{AB}]=\vartheta[P''''_{AB}]=\vartheta[{\mathcal{T}_{d+1}}_AI_BP'''_{AB}]\leq
\vartheta[P'''_{AB}]$.

\medskip

Using Lemmas 4 and 5, and noting that ${\mathcal{K}^{(2)}_{d+1}}_A$
and ${\mathcal{K}^{(3)}}_A$ are reversible, we obtain
$\vartheta[\mathcal{\bar D}_A{\bar \G}_B\bar
P_{AB}]=\vartheta[{\mathcal{W}_{{d_A}+1}}_AI_B\,
P^{''}_{AB}]\leq\vartheta[P'''_{AB}]\leq\vartheta[P''_{AB}]$.

\medskip

From Lemma 6 and the similarity of $\mathcal G_c$ to  $\mathcal G'$
we find that $\vartheta[P''_{AB}]=
\vartheta[{\mathcal{G}_{{d_A}+1}}_AI_BP^{'}_{AB}]\leq
\vartheta[P^{'}_{AB}]$. It follows that $\vartheta[\mathcal{\bar
D}_A{\bar \G}_B\bar P_{AB}]\leq \vartheta[P^{'}_{AB}]$.

\medskip

If we look at the form of $P^{'}_{AB}$ we find that the same
decomposition can be performed on the operations
${\mathcal{W}_{{d_A}}}_AI_B {\mathcal{G}_{{d_A}}}_AI_B $. It is
straightforward to use the above arguments to show that
$\vartheta[P'_{AB}]=\vartheta[{\mathcal{W}_{{d_A}}}_AI_B
{\mathcal{G}_{{d_A}}}_AI_BP^y_{AB}]\leq\vartheta[P^y_{AB}]$.

\medskip

By repeated use of the above arguments and a study of Eq.
(\ref{biffy}) one finds that $\vartheta[\mathcal{\bar D}_A{\bar
\G}_B\bar P_{AB}]\leq\vartheta[\mathcal \R_A \I_B \;\;I_A{\bar
\G}_B\bar P_{AB}]$. Since $\R_A$ is
 reversible, by Lemma 4, $\vartheta[\mathcal \R_A \I_B\,
\I_A{\bar \G}_B\bar P_{AB}]=\vartheta[\I_A\mathcal {\bar D}_B\bar
 P_{AB}]$.

\medskip

 Exactly the same arguments can be used to show that $\vartheta[\I_A\mathcal {\G}_B\bar
 P_{AB}]\leq \vartheta[\bar
 P_{AB}]$. It follows that $\vartheta[\mathcal{\bar D}_A{\bar \G}_B\bar
P_{AB}]\leq\vartheta[\bar P_{AB}]$. $\blacksquare$


Noting the definition of the function $\vartheta$ and $\bar P_{AB}$
Eq. (\ref{l0y}) follows.

\end{document}